\documentclass[aps,prd,preprintnumbers,amsmath,amssymb,nofootinbib,11pt]{revtex4}
\usepackage{eepic}
\usepackage{indentfirst}
\usepackage{mathrsfs}
\usepackage{fancyhdr}
\usepackage{ulem}
\usepackage{float}
\usepackage{graphicx}
\usepackage[
colorlinks=true, linkcolor=black, breaklinks=true, urlcolor=blue,
citecolor=green]{hyperref}
\usepackage{epstopdf}
\usepackage{bm,bbm}

\usepackage{xcolor}

\usepackage{tabularx}
\usepackage{subfigure}
\usepackage{epsfig}
\usepackage{color}
\usepackage{slashed}
\usepackage{hyperref}

\newcommand{\be}{\begin{equation}}
\newcommand{\ee}{\end{equation}}
\renewcommand{\L}{\mathscr{L}}

\newcommand{\bra}{\langle}
\newcommand{\ket}{\rangle}

\newcommand{\MeV}{\,\text{MeV}}
\newcommand{\GeV}{\,\text{GeV}}
\renewcommand{\vec}[1]{\mathbf{#1}}
\newcommand{\diff }{{\text{d}}}

\begin{document}
\pagestyle{plain}

\title {\boldmath Study of the $Y(4660)$ from a light-quark perspective}
\author{ Yun-Hua~Chen}\email{yhchen@ustb.edu.cn}
\affiliation{School of Mathematics and Physics, University
of Science and Technology Beijing, Beijing 100083, China}

\begin{abstract}

In this paper, we try to reveal the structure of the $Y(4660)$ from the light-quark perspective.
We study the dipion invariant mass spectrum and the helicity angular distributions of the  $e^+ e^-
\to Y(4660) \to \psi(2S) \pi^+\pi^-$ process. In
particular, we consider the effects of different light-quark SU(3)
eigenstates inside the $Y(4660)$. The strong pion-pion final-state interactions
as well as the $K\bar{K}$ coupled channel in the $S$-wave
are taken into account model independently by using dispersion
theory.
We find that the light-quark SU(3) octet state plays a significant
role in this transition, implying that the $Y(4660)$ contains a
large light-quark component and thus might not be a pure conventional charmonium state.
In the fit scheme considering both the SU(3) singlet and SU(3) octet states, two solutions are found,
and both solutions reproduce the $\pi\pi$
invariant mass spectra well. New measurement data with higher statistics in the future will be helpful to better distinguish these two solutions.

\end{abstract}

\maketitle

\newpage
\section{Introduction}

In recent years, a number of charmoniumlike states have been discovered and they challenge our
current understanding of hadron spectroscopy. Among these states, the $Y(4660)$ was first observed in the
initial-state radiation process
$e^+e^-\to \gamma_{ISR} \psi(2S)\pi^+\pi^-$ by the Belle Collaboration~\cite{Wang:2007ea}.
There is no charmonium state expected in the $Y(4660)$ mass region with quantum numbers $1^{--}$ from the naive quark model~\cite{Godfrey}, and the $Y(4660)$ was not observed in $e^+e^-\to \gamma_{ISR} J/\psi\pi^+\pi^-$.
Such peculiar properties have initiated a
lot of theoretical and experimental studies, see Refs.~\cite{Chen:2016qju,Hosaka:2016pey,Lebed:2016hpi,Esposito:2016noz,Guo:2017jvc,Ali:2017jda,Olsen:2017bmm,Karliner:2017qhf,Yuan:2018inv,Kou:2018nap,Cerri:2018ypt,Guo:2019twa} for recent reviews. On the theoretical side, the
$Y(4660)$ has been interpreted as an excited
charmonium~\cite{Ding:2007rg,Li:2009zu,Wang:2020prx,Wang:2020kej}, a hadronic molecule of
$\psi(2S)f_0(980)$ or $\Lambda_c \bar{\Lambda}_c$~\cite{Guo:2008zg,Guo:2010tk,Dai:2017fwx}, a tetraquark state with diquark-antidiquark [$cs$][$\bar{c}\bar{s}$] and [$cq$][$\bar{c}\bar{q}$] type \cite{Ebert:2008kb,Albuquerque:2008up,Albuquerque:2011ix,Chen:2010ze,Zhang:2010mw,Sundu:2018toi,Wang:2019iaa,Wang:2018rfw}, a charmed baryonium\cite{Qiao:2007ce,Cotugno:2009ys}, and
a hadrocharmonium~\cite{Dubynskiy:2008mq}. On the experimental side, the signals of the states
in different channels such as
$e^+e^-\to \psi(2S)\pi^+\pi^-$~\cite{Wang:2007ea,Wang:2014hta},
$\Lambda_c \bar{\Lambda}_c$~\cite{Pakhlova:2008vn},
and $D_s^+ D_{s1}(2536)^{-}+\mathrm{c.c.}$~\cite{Jia:2019gfe}
have been analyzed and attributed to the $Y(4660)$, as adopted in PDG.~\cite{Zyla:2020zbs}.
Note that very recently, the BESIII collaboration reported a charged hidden-charm structure with strangeness, which is named $Z_{cs}(3985)$, in the
$e^+e^-\to K^+(D_s^- D^{*0}+D_s^{*-}D^0)$ process~\cite{Ablikim:2020hsk}. The measurement indicated that a clear signal of $Z_{cs}(3985)$ only appears at the
center of mass energy of 4.681 GeV, in the vicinity of the $Y(4660)$.

In the present work, we will study the possible light-quark components of
the $Y(4660)$ to help reveal its internal structure. We will focus
on the $\pi\pi$ invariant mass spectrum of the reaction
$e^+e^-\to Y(4660) \to \psi(2S) \pi\pi$, which was presented after applying an appropriate cut
to the $\psi(2S) \pi\pi$ invariant mass in Ref.~\cite{Wang:2014hta}. In this process, the $\pi\pi$ invariant mass
can reaches above the $K\bar{K}$ threshold, and thus allows us to extract the
information of the light-quark SU(3) flavor-singlet and flavor-octet components.
If the $Y(4660)$
contains no light quarks (as in the charmonium
scenario), the light-quark source provided by the $Y(4660)$ has
to be in the form of an SU(3) singlet state. Therefore the determination
of the contributions from different SU(3) eigenstate components is
instructive to clarify the internal structure of the $Y(4660)$, especially in
the case if a nonzero SU(3) octet component is found to be
indispensable to reproduce the experimental data. The similar strategy has been applied to study the nature of the
$Y(4260)$ state in our previous work~\cite{Chen:2019mgp}.
The main difference is that in the present work we simultaneously
fit to the experimental data of the $\pi\pi$ invariant mass distributions and the helicity angular distributions, while in Ref.~\cite{Chen:2019mgp} the helicity angular distribution data were not considered.

\begin{figure}
\centering
\includegraphics[width=\linewidth]{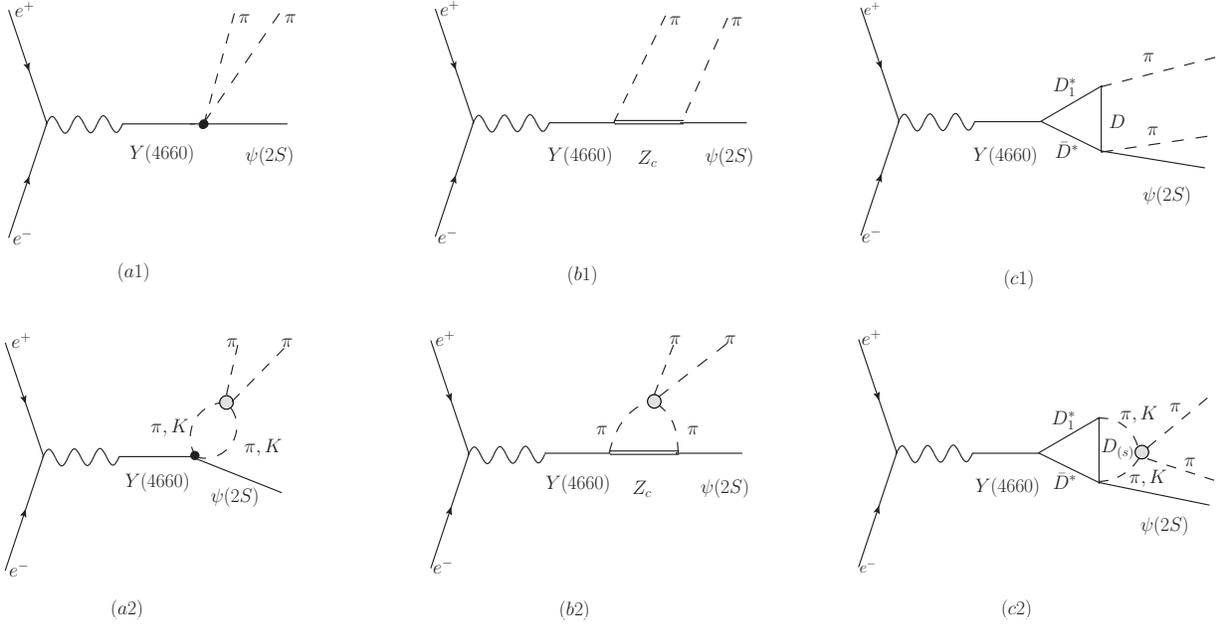}
\caption{Feynman diagrams considered for $e^+ e^- \to
Y(4660) \to \psi(2S) \pi \pi $. The gray blob denotes the effects of FSI.
}\label{fig.FeynmanDiagram}
\end{figure}

Parity and $C$-parity conservation require the dipion system in $e^+e^-\to Y(4660) \to \psi(2S) \pi\pi$
to be in even partial waves. The dipion invariant mass can reaches above the $K\bar{K}$ threshold, so the coupled-channel final-state interactions (FSIs) in the $S$-wave is strong and needs to be taken into account.
Based on unitarity and
analyticity, the modified Omn\`es representation
is used in this study, where the left-hand-cut contribution is approximated
by the sum of the $Z$-exchange mechanism and the triangle diagrams $Y(4660) \rightarrow \bar{D}^\ast D_1^\ast(2600)\rightarrow \bar{D}^\ast D \pi (\bar{D}^\ast D_s K) \rightarrow \psi(2S)\pi\pi (\psi(2S)K\bar{K})$.\footnote{We also need
to take account of the $ Y(4660) \to \psi(2S)
K \bar K $  process in the coupled-channel FSI.}
\footnote{Note that no $Z$ structure is observed in the $\psi(2S)\pi$ channel in the $e^+ e^-
\to Y(4660) \to \psi(2S) \pi^+\pi^-$ process, and the reason may be that the present experimental data is limited in statistics~\cite{Wang:2014hta}.
Given the significant role of the intermediate $Z_c(3900)$ plays in the $Y(4260) \to J/\psi \pi^+\pi^-$ process, the contribution mediated by the flavor partner may
give some important effects in the $Y(4660) \to \psi(2S) \pi^+\pi^-$ transition. Here we take account of the possible effect of the $Z_c(4430)$-exchange, since the $Z_c(4430)$ is the only $Z$ state with the decay mode of $\psi(2S)\pi$ observed according to PDG. A better distinction of the effect of the $Z$ state requires a detailed analysis of the $\psi(2S)\pi$ distribution data with higher statistics in the future.}
At low energies, the amplitude should agree with the leading chiral contact results.
For the leading contact couplings for $Y(4660)\psi(2S)\pi\pi$ and $Y(4660)\psi(2S) K\bar K$, we construct the chiral Lagrangians in the spirit of
the chiral effective field theory ($\chi$EFT) and the heavy-quark nonrelativistic
expansion~\cite{Mannel}. The parameters are then determined from fitting to the Belle data.
The relevant Feynman diagrams considered are given in Fig.~\ref{fig.FeynmanDiagram}.

This paper is organized as follows. In Sec.~\ref{theor}, we introduce
the theoretical framework and elaborate on the calculation of the
transition amplitudes as well as the dispersive treatment of the FSI. In Sec.~\ref{pheno}, we present the fit results
and discuss the light-quark components of the $Y(4660)$ and its
structure. A summary is given in Sec.~\ref{conclu}.

\section{Theoretical framework}\label{theor}
\subsection{Lagrangians}

In general, the light-quark sources provided by the $Y(4660)$ in the $Y(4660) \to \psi(2S) \pi^+\pi^-$ transition may come from two ways, one is from the possible light-quark components contained in the $Y(4660)$ (e.g., in the charm- and anticharm-mesons molecule or the four-quark scenarios), and the other way is that the light quarks are excited by the $Y(4660)$ from vacuum (e.g., in the pure $c \bar{c}$ or the hybrid state scenarios). Here we do not distinguish these two types of possible light-quark sources, but take them into account in an unified scheme, since what is matter here is the relative strengths between the light-flavor SU(3) singlet part and SU(3) octet part acting in this transition. Considering the light-quarks sources provided by the $Y(4660)$, the $Y(4660)$ can be decomposed into SU(3) singlet and octet components of light quarks,
\be
\label{eq.YComponents} |Y(4660)\rangle=a|V_1\rangle+b|V_8\rangle\,,
\ee
where $|V_1\rangle \equiv V_1^{\text{light}}\otimes
V^{\text{heavy}}=
\frac{1}{\sqrt{3}}(u\bar{u}+d\bar{d}+s\bar{s})\otimes
V^{\text{heavy}}$ and $|V_8\rangle \equiv
V_8^{\text{light}}\otimes
V^{\text{heavy}}=\frac{1}{\sqrt{6}}(u\bar{u}+d\bar{d}-2
s\bar{s})\otimes V^{\text{heavy}}$.
Note that the heavy-quark (e.g., $c$ quark) components are contained in the $V^{\text{heavy}}$, and they are not
distinguishable in $|V_1\rangle$ and $|V_8\rangle$.
Expressed in terms of a $3\times3$ matrix in the SU(3) flavor space, the $Y(4660)$ is written as
\begin{equation}
 \frac{a}{\sqrt{3}} V_1 \cdot \mathbbm{1}+\frac{b}{\sqrt{6}} V_8\cdot \text{diag} \left( 1,  1, - 2\right)    .
\end{equation}

The effective Lagrangian for the $Y(4660)\psi(2S)\pi\pi$ and
$Y(4660)\psi(2S) K\bar{K}$ contact couplings, at leading order in the chiral as well as the heavy-quark nonrelativistic expansion,
reads
~\cite{Mannel,Chen:2019mgp}
\begin{equation}\label{LagrangianYpsi2Spipi}
\L_{Y\psi^\prime\Phi\Phi} = g_1\bra V_{1}^\alpha J^\dag_\alpha \ket \bra u_\mu
u^\mu\ket +h_1\bra V_{1}^{\alpha} J^\dag_\alpha \ket \bra u_\mu u_\nu\ket
v^\mu v^\nu +g_8\bra  J^\dag_\alpha \ket \bra V_{8}^{\alpha} u_\mu u^\mu\ket
+h_8\bra J^\dag_\alpha \ket \bra V_{8}^{\alpha} u_\mu u_\nu\ket v^\mu v^\nu
+\mathrm{H.c.}\,,
\end{equation}
where $\langle\ldots\rangle$ denotes the trace in the SU(3) flavor space, $J= (\psi^\prime/\sqrt{3}) \cdot \mathbbm{1}$, and
$v^\mu=(1,\vec{0})$ is the velocity of the heavy quark. The lightest pseudoscalar mesons can
be filled nonlinearly into
\begin{equation}
u_\mu = i \left( u^\dagger\partial_\mu u\, -\, u \partial_\mu
u^\dagger\right) \,, \qquad
u = \exp \Big( \frac{i\Phi}{\sqrt{2}F} \Big)\,,
\end{equation}
with the Goldstone fields
\begin{align}
\Phi &=
 \begin{pmatrix}
   {\frac{1}{\sqrt{2}}\pi ^0 +\frac{1}{\sqrt{6}}\eta _8 } & {\pi^+ } & {K^+ }  \\
   {\pi^- } & {-\frac{1}{\sqrt{2}}\pi ^0 +\frac{1}{\sqrt{6}}\eta _8} & {K^0 }  \\
   { K^-} & {\bar{K}^0 } & {-\frac{2}{\sqrt{6}}\eta_8 }  \\
 \end{pmatrix} . \label{eq:u-phi-def}
\end{align}
Here $F$ corresponds to the pion decay constant, $F_\pi=92.1\MeV$, in the chiral limit.

We need the $Z_c Y(4660)\pi$ and the $Z_c \psi(2S)
\pi$ interacting Lagrangians to calculate the contribution of the
intermediate $Z_c$ states, namely $Y(4660) \to Z_c\pi \to \psi(2S) \pi\pi$.
The leading-order Lagrangians are~\cite{Guo2011}
\begin{align}
\L_{Z_c Y\pi} = C_{Z_c Y\pi}  Y^i \bra {Z^i_{c}}^\dagger u_\mu \ket v^\mu
+\mathrm{H.c.} \,, \notag\\
\L_{Z_c\psi^\prime\pi} = C_{Z_c \psi^\prime\pi} {\psi^\prime}^i \bra {Z^i_{c}}^\dagger u_\mu \ket
v^\mu +\mathrm{H.c.} \,. \label{LagrangianZc}
\end{align}

In order to calculate the triangle diagrams $Y(4660) \rightarrow \bar{D}^\ast D_1^\ast(2600)\rightarrow \bar{D}^\ast D \pi (\bar{D}^\ast D_s K) \rightarrow \psi(2S)\pi\pi (\psi(2S)K\bar{K})$,\footnote{Here and in the following, $\bar{D}^\ast D_1^\ast$ always means the negative $C$-parity combination of $\bar{D}^\ast D_1^\ast$ and $D^\ast \bar{D}_1^\ast$.}
we need the Lagrangians for the coupling of the $Y(4660)$ to $\bar{D}^\ast D_1^\ast$ as well as the couplings of the $D_1^\ast$ to $D\pi$ and $D_s K$,
\begin{align}\nonumber
\L_{Y D_1^* D^*}=& g_{Y D_1^* D^*} \big[-Y^\mu D_1^{*\nu}\overleftrightarrow{\partial}_\mu D_\nu^{*\dagger}
+ Y^\mu D_1^{*\nu}\partial_\nu{D}^{*\dagger}_{\mu} -
Y^\mu\partial_\nu D_1^{*\nu}{D}_\mu^{*\dagger}
+ Y^\mu D^{*\nu\dagger}\partial_\nu{D_1}^{*}_{\mu} \\&-
Y^\mu\partial_\nu D^{*\nu\dagger}D_{1\mu}^{*}\big]+{\rm H.c.} \,, \\
 \L_{D_1^* D P}=& i\frac{g_{D_1^* D P}}{F}
\big[D_{1\mu}^*\partial^\mu\pi \bar{D}-D\partial^{\mu}\pi \bar{D}_{1\mu}^*
+D_{1\mu}^*\partial^\mu K \bar{D}-D\partial^{\mu}K \bar{D}_{1\mu}^* \big]+{\rm H.c.} \,,
\label{LagrangianD1}
\end{align}
where $P$ denotes the pseudoscalar meson $\pi$ or $K$.
We also need the Lagrangian for the $\psi(2S) D^\ast D \pi$ vertex, which at leading order in heavy-meson chiral perturbation theory
is
\begin{equation}\label{Lagrangianpsi2SDstarDpi}
\L_{\psi^\prime D^\ast D P}= \frac{g_{\psi^\prime P}}{2} \bra \psi^\prime\bar{H}_a^\dagger H_b^\dagger  \ket u_{ab}^0\,,
\end{equation}
where the charm mesons are collected in
$H_a=\vec{V}_a \cdot \boldsymbol{\sigma}+P_a$ with
$P_a(V_a)=(D^{(*)0},D^{(*)+},D_s^{(*)+})$, and
$\bar{H}_a=- \bar{\vec{V}}_a \cdot \boldsymbol{\sigma}+\bar{P}_a$
with
$\bar{P}_a(\bar{V}_a)=(\bar{D}^{(*)0},D^{(*)-},D_s^{(*)-})$~\cite{Mehen2008}.

The gauge-invariant $\gamma^\ast(\mu)$ and $Y(4660) (\nu)$ interaction can be written as
\be
iV_{\gamma^{\ast\mu}Y^\nu}=2i(g^{\mu\nu}p^2-p^\mu
p^\nu)c_\gamma \,,
\ee
where $p$ is the momentum of the virtual photon $\gamma^\ast.$

\subsection{Amplitudes of \boldmath{$ Y(4660) \to \psi(2S) PP $} processes}

The decay amplitude of $ Y(4660)(p_a) \to
\psi^\prime(p_b) P(p_c)P(p_d) $ can be described in terms of the
Mandelstam variables
\begin{align}
s &= (p_c+p_d)^2 , \qquad
t_P=(p_a-p_c)^2\,, \qquad u_P=(p_a-p_d)^2\,.
\end{align}
We define $\vec{q}$ as the
three-momentum of the final $\psi(2S)$ in the rest frame of the $Y(4660)$ with
\be \label{eq:q} |\vec{q}|=\frac{1}{2M_{Y}}
\lambda^{1/2}\big(M_{Y}^2,M_{\psi^\prime}^2,s\big) \,, \ee
where
$\lambda(a,b,c)=a^2+b^2+c^2-2(ab+ac+bc)$ is the K\"all\'en triangle function.

Using the Lagrangians in
Eq.~\eqref{LagrangianYpsi2Spipi}, the projections of the $S$- and $D$-waves of the chiral contact terms are obtained as
\begin{align}
M_0^{\chi,\pi}(s)&=-\frac{2}{F_\pi^2}\sqrt{M_{Y}M_{\psi^\prime}}
\bigg\{\Big(g_1+\frac{g_8}{\sqrt{2}}\Big) \left(s-2m_\pi^2 \right)
+\frac{1}{2}\Big(h_1+\frac{h_8}{\sqrt{2}}\Big) \bigg[s+\vec{q}^2\Big(1
-\frac{\sigma_\pi^2}{3} \Big)\bigg]\bigg\}\,, \notag\\
M_0^{\chi,K}(s)&=-\frac{2}{F_K^2}\sqrt{M_{Y}M_{\psi^\prime}}
\bigg\{\Big(g_1-\frac{g_8}{2\sqrt{2}}\Big) \left(s-2m_K^2 \right)
+\frac{1}{2}\Big(h_1-\frac{h_8}{2\sqrt{2}}\Big) \bigg[s+\vec{q}^2\Big(1
-\frac{\sigma_K^2}{3} \Big)\bigg]\bigg\}\,, \notag\\
\label{eq.M0+2Pi+Kchiral}
M_2^{\chi,\pi}(s)&=\frac{2}{3
F_\pi^2}\sqrt{M_{Y}M_{\psi^\prime}}\,\Big(h_1+\frac{h_8}{\sqrt{2}}\Big)
|\vec{q}|^2\sigma_\pi^2\,,
\end{align}
where the kaon decay constant $F_K=0.113$ GeV has been employed, and $\sigma_P \equiv
\sqrt{1-4m_P^2/s}$.
For the $D$-wave, the single-channel FSI will be taken into account and we only
give the amplitude of the process involving pions.

For the $Y(4660) \to \psi(2S) \pi^+ \pi^-$ process, since the
crossed-channel exchanged $Z$ and $DD^\ast$ can be on-shell, the left-hand cut (l.h.c.)
produced intersects and overlaps with the right-hand cut (r.h.c.) and requires special treatment.
As discussed in Ref.~\cite{Schmid:1967ojm}, the l.h.c.\ is in fact in the unphysical Riemann sheet. The proper analytical continuation for the energy variable $q^2 \to q^2+i\epsilon$ helps to locate the l.h.c.\ in the right position so that it does not overlap with the r.h.c.\ in the physical Riemann sheet. Also we will take into account the finite width in the $D_1^*$ propagator.

Using the Lagrangians in
Eq.~\eqref{LagrangianZc}, the projections of $S$- and $D$-waves
of the $Z_c$-exchange amplitude are obtained as
\begin{align}\label{eq.M0M2Zc}
\hat{M}_0^{Z_c,\pi}(s)&=-\frac{2 \sqrt{M_Y M_{\psi^\prime}}M_{Z_c}}{ F^2
\kappa_\pi(s)}C_{Y\psi^\prime}^{Z_c}\Big\{
\big(s+|\vec{q}|^2\big)Q_0(y(s))
 -|\vec{q}|^2\sigma_\pi^2\big[y^2(s)
Q_0(y(s))-y(s)\big] \Big\},\nonumber\\
\hat{M}_2^{Z_c,\pi}(s)&=-\frac{5 \sqrt{M_Y M_{\psi^\prime}}M_{Z_c}}{ F^2
\kappa_\pi(s)}C_{Y\psi^\prime}^{Z_c}\Big\{ \big[s+|\vec{q}|^2-|\vec{q}|^2\sigma_\pi^2
y^2(s)\big]
\times\big[(3y^2(s)-1)Q_0(y(s))-3y(s)\big]
\Big\}\,,
\end{align}
where $\kappa_\pi(s) \equiv \sigma_\pi
\lambda^{1/2}(M_{Y}^2,M_{\psi^\prime}^2,s)$, $C_{Y\psi^\prime}^{Z_c}\equiv C_{Z_c Y\pi} C_{Z_c \psi^\prime\pi}$ is the product
of the coupling constants for the exchange of the $Z_c$, $y(s)\equiv {(3s_0-s-2 M_{Z_c}^2+2 iM_{Z_c}\Gamma_{Z_c})}/{\kappa_\pi(s)}$,
and $Q_0(y)$ is the Legendre function of the second kind,
\begin{equation}\label{eq.Q0}
Q_0(y)=\frac{1}{2}\int_{-1}^1 \frac{\diff z}{y-z}P_0(z)
 = \frac{1}{2}\log \frac{y+1}{y-1}\,.
\end{equation}

In the calculation of the triangle diagrams, we only keep the terms proportional to
$\bm{\epsilon}_{Y}\cdot \bm{\epsilon}_{\psi}$, and omit the remaining terms proportional to contractions of
momenta with the polarization vectors, which are suppressed in the heavy-quark nonrelativistic expansion~\cite{Chen:2016mjn}.
Explicitly, the partial-wave projections of the triangle amplitude for the $ Y(4660) \to \psi(2S) \pi\pi (\psi(2S) K \bar{K}) $ process read
\begin{align}\label{eq.MlLoop}
\hat{M}_l^{\text{loop},\pi(K)}(s)&=\frac{2l+1}{2}\frac{8\sqrt{2} \sqrt{M_Y M_\psi}M_{D_1^*}M_{D}M_{D_{(s)}^\ast}}{ F_{\pi(K)}^2
}C_{Y\psi^\prime}^{\text{loop}}\int_{-1}^1 \diff \cos\theta P_l(\cos\theta)
\nonumber\\
&\times  \int  \frac{\diff^d l}{(2\pi)^d}\bigg\{\frac{i (\vec{p_a}-2\vec{l})\cdot\vec{p_d} p_c^0}{(l^2-M_{D_1^*}^2+i M_{D_1^*} \Gamma_{D_1^*}+i\epsilon)\big[(p_a-l)^2-M_{D^*}^2+i\epsilon\big]\big[(l-p_d)^2-M_{D}^2+i\epsilon\big]}
\nonumber\\& \qquad +\frac{i (\vec{p_a}-2\vec{l})\cdot\vec{p_c} p_d^0}{(l^2-M_{D_1^*}^2+i M_{D_1^*} \Gamma_{D_1^*}+i\epsilon)\big[(p_a-l)^2-M_{D^*}^2+i\epsilon\big]\big[(l-p_c)^2-M_{D}^2+i\epsilon\big]}\bigg\}\,,
\end{align}
where $C_{Y\psi^\prime}^{\text{loop}}\equiv g_{Y D_1^*D^*} g_{D_1^* D P} g_{\psi^\prime P}$ is the product of the coupling constants for the triangle diagrams.

\subsection{Final-state interactions with a dispersive approach, Omn\`es solution }

There are strong FSIs in the $\pi\pi$ system, which can be taken into account
model-independently using dispersion theory. Based on unitarity and
analyticity, the Omn\`es solutions
will be used in this study.
Similar methods to consider the FSI have been applied previously e.g.\ in
Refs.~\cite{Moussallam-gamma,KubisPlenter,ZHGuo,Kang,Dai:2014lza,Dai:2014zta,Dai:2016ytz,Chen2016,Chen:2016mjn,Chen:2019gty,Chen:2019mgp}.
Because the invariant
mass of the pion pair reaches above the $K\bar{K}$ threshold, we
will take account of the coupled-channel ($\pi\pi$ and $K\bar
K$) FSIs for the dominant $S$-wave component, while for the $D$-wave
only the single-channel ($\pi\pi$) FSI will be considered.

For $Y(4660) \to \psi(2S) \pi^+ \pi^-$, the
partial-wave decomposition of the amplitude including the $s$-channel FSI reads
\be
M^\text{decay}(s,\cos\theta)  = \sum_{l=0}^{\infty}
\left[M_l^\pi(s)+\hat{M}_l^\pi(s)\right] P_l(\cos\theta)\,,
\label{eq.PartialWaveFullAmplitude}\ee
where $M_l^\pi(s)$ includes
the r.h.c.\ part and accounts for the $s$-channel rescattering,
and the ``hat function'' $\hat{M}_l^\pi(s)$ contains the l.h.c., contributed by the possible crossed-channel pole terms or the open-flavor loop effects.
In this study, we approximate the l.h.c.\ by the sum of the $Z_c$-exchange diagram and the triangle diagrams,
i.e., $\hat{M}_l^\pi(s)=\hat{M}_l^{Z_c,\pi}(s)+\hat{M}_l^{\text{loop},\pi}(s)$. $\theta$ is the angle between the positive
pseudoscalar meson and the $Y(4660)$ in the rest frame of the $PP$ system.

For the $S$-wave, we will take into account the two-channel
rescattering effects. The two-channel unitarity condition reads
\begin{equation}\label{eq.unitarity2channel}
\textrm{disc}\, \vec{M}_0(s)=2i T_0^{0\ast}(s)\Sigma(s)
\left[\vec{M}_0(s)+\hat{\vec{M}}_0(s)\right] ,
\end{equation}
where the two-dimensional vectors $\vec{M}_0(s)$ and
$\hat{\vec{M}}_0(s)$ represent the r.h.c.\ and the l.h.c.\
parts of both the $\pi\pi$ and the $K\bar{K}$ final states,
respectively,
 \begin{equation}
\vec{M}_0(s)=\left( {\begin{array}{*{2}c}
   {M^\pi_0(s)} \\
   {\frac{2}{\sqrt{3}}M^K_0(s)}  \\\end{array}} \right), \hspace{0.5cm}\hat{\vec{M}}_0(s)=\left( {\begin{array}{*{2}c}
   {\hat{M}^\pi_0(s)} \\
   {\frac{2}{\sqrt{3}}\hat{M}^K_0(s)}  \\
\end{array}} \right).
 \end{equation}

The two-dimensional matrices $T_0^0(s)$ and $\Sigma(s)$ are given by
\begin{equation}\label{eq.T00}
T_0^0(s)=
 \left( {\begin{array}{*{2}c}
   \frac{\eta_0^0(s)e^{2i\delta_0^0(s)}-1}{2i\sigma_\pi(s)} & |g_0^0(s)|e^{i\psi_0^0(s)}   \\
  |g_0^0(s)|e^{i\psi_0^0(s)} & \frac{\eta_0^0(s)e^{2i\left(\psi_0^0(s)-\delta_0^0(s)\right)}-1}{2i\sigma_K(s)} \\
\end{array}} \right),
\end{equation}
and $\Sigma(s)\equiv \text{diag}
\big(\sigma_\pi(s)\theta(s-4m_\pi^2),\sigma_K(s)\theta(s-4m_K^2)\big)$.
There are three input functions entering the $T_0^0(s)$ matrix: the
$\pi\pi$ isoscalar $S$-wave phase shift $\delta_0^0(s)$, and the modulus and phase of the $\pi\pi \to
K\bar{K}$ $S$-wave amplitude $g_0^0(s)=|g_0^0(s)|e^{i\psi_0^0(s)}$. We will use the parametrization of the $T_0^0(s)$ matrices given in Refs.~\cite{Dai:2014lza,Dai:2014zta,Dai:2016ytz}.
Note that the inelasticity $\eta_0^0(s)$ in Eq.~\eqref{eq.T00} is
related to the modulus $|g_0^0(s)|$ as
\begin{equation}
\eta_0^0(s)=\sqrt{1-4\sigma_\pi(s)\sigma_K(s)|g_0^0(s)|^2\theta(s-4m_K^2)}\,.
\end{equation}
These inputs are used up to $\sqrt{s_0}=1.3\GeV$, below the onset of
further inelasticities from the $f_0(1370)$ and $f_0(1500)$ resonances which couple strongly to
$4\pi$~\cite{Tanabashi:2018oca,Ropertz:2018stk}. Above $s_0$, the phases
$\delta_0^0(s)$ and $\psi_0^0$ are guided smoothly to 2$\pi$~\cite{Moussallam2000}
\begin{equation}
\delta(s)=2\pi+(\delta(s_0)-2\pi)\frac{2}{1+({s}/{s_0})^{3/2}}\,.
\end{equation}

The solution of the unitarity condition in
Eq.~\eqref{eq.unitarity2channel} is given by
\begin{equation}\label{OmnesSolution2channel}
\vec{M}_0(s)=\Omega(s)\bigg\{\vec{P}^{n-1}(s)+\frac{s^n}{\pi}\int_{4m_\pi^2}^\infty
\frac{\diff
x}{x^n}\frac{\Omega^{-1}(x)T(x)\Sigma(x)\hat{\vec{M}}_0(x)}{x-s}\bigg\}
\,,
\end{equation}
where $\Omega(s)$ satisfies the homogeneous coupled-channel
unitarity relation
\begin{equation}\label{eq.unitarity2channelhomo}
\textrm{Im}\, \Omega(s)=T_0^{0\ast}(s)\Sigma(s) \Omega(s),
\hspace{1cm}  \Omega(0)=\mathbbm{1} \,,
\end{equation}
and its numerical results have been computed, e.g., in
Refs.~\cite{Leutwyler90,Moussallam2000,Hoferichter:2012wf,Daub}.

For the $D$-wave, we will take account of the single-channel FSI. In the elastic $\pi\pi$ rescattering region,
the partial-wave unitarity condition reads
\begin{equation}\label{eq.unitarity1channel}
\textrm{Im}\, M_2(s)= \left[M_2(s)+\hat{M}_2(s)\right]
\sin\delta_2^0(s) e^{-i\delta_2^0(s)}\,,
\end{equation}
where the phase of the isoscalar $D$-wave amplitude
$\delta_2^0$ coincides with
the $\pi\pi$ elastic phase shift, as required by
Watson's theorem~\cite{Watson1,Watson2}.
The modified Omn\`es solution of Eq.~\eqref{eq.unitarity1channel}
is~\cite{Leutwyler96,Chen2016}
\be\label{OmnesSolution1channel}
M_2(s)=\Omega_2^0(s)\bigg\{P_2^{n-1}(s)+\frac{s^n}{\pi}\int_{4m_\pi^2}^\infty
\frac{\diff x}{x^n} \frac{\hat
M_2(x)\sin\delta_2^0(x)}{|\Omega_2^0(x)|(x-s)}\bigg\} \,, \ee
where the polynomial $P_2^{n-1}(s)$ is a subtraction function, and the
Omn\`es function is defined as~\cite{Omnes}
\begin{equation}\label{Omnesrepresentation}
\Omega_2^0(s)=\exp
\bigg\{\frac{s}{\pi}\int^\infty_{4m_\pi^2}\frac{\diff x}{x}
\frac{\delta_2^0(x)}{x-s}\bigg\}\,.
\end{equation}
We will use the Madrid--Krak\'ow group~\cite{Pelaez} result for $\delta_2^0(s)$, which is smoothly continued to $\pi$ for $s\to\infty$.

On the other hand, at
low energies the amplitudes $\vec{M}_0(s)$ and $M_2(s)$ should match to those from $\chi$EFT. Namely, in the limit of
switching off the FSI at $s=0$, $\Omega(0)=\mathbbm{1}$, and $\Omega_2^0(0)=1$,
the subtraction terms should agree well with the low-energy chiral results
given in Eq.~\eqref{eq.M0+2Pi+Kchiral}.
Therefore, for the
$S$-wave, the amplitude takes the form
\begin{equation}\label{M02channel}
\vec{M}_0(s)=\Omega(s)\bigg\{\vec{M}_0^{\chi}(s)+\frac{s^3}{\pi}\int_{4m_\pi^2}^\infty
\frac{\diff
x}{x^3}\frac{\Omega^{-1}(x)T(x)\Sigma(x)\hat{\vec{M}}_0(x)}{x-s}\bigg\}
\,,
\end{equation}
where $ \vec{M}^{\chi}_0(s)=\big(
   M_0^{\chi,\pi}(s),
   2/\sqrt{3}\,M_0^{\chi,K}(s)
   \big)^{T}$, while
for the $D$-wave, it can be written as
\be\label{M21channel}
M_2^\pi(s)=\Omega_2^0(s)\bigg\{M_2^{\chi,\pi}(s)+\frac{s^3}{\pi}\int_{4m_\pi^2}^\infty
\frac{\diff x}{x^3} \frac{\hat
M_2^\pi(x)\sin\delta_2^0(x)}{|\Omega_2^0(x)|(x-s)}\bigg\} \,.
\ee

The polarization-averaged modulus square of the $e^+e^- \to Y(4660)
\to \psi(2S) \pi^+\pi^-$ amplitude can be written as
\begin{equation}
|\bar{M}(E^2,s,\cos\theta)|^2 = \frac{4\pi\alpha
c_\gamma^2|M^{\text{decay}}(s,\cos\theta)|^2}{3|E^2-M_Y^2+iM_Y\Gamma_Y|^2
M_{\psi^\prime}^2}\left[  8 M_{\psi^\prime}^2 E^2+(s-E^2-M_{\psi^\prime}^2)^2
\right],\label{eq.eetoJpsipipiAmplitudeSquar}
\end{equation}
where $E$ is the center of mass energy of the
$e^+e^-$ collisions, and we set the $\gamma^\ast Y(4660)$ coupling
constant $c_\gamma$ to 1 since it can be absorbed into the overall
normalization in the fit to the event distributions. Here we use the energy-independent width for the
$Y(4660)$, and the values of the $Y(4660)$ mass and width are taken as $4633\MeV$ and $64.0\MeV$, respectively, which are the central values in PDG.~\cite{Zyla:2020zbs}. We also have tried to allow the
mass and width of the $Y(4660)$ to float freely, and found
that the fit quality changes only slightly.
At last, the $\pi\pi$ invariant mass spectra and the helicity angular distribution for $e^+e^- \to \psi(2S) \pi^+\pi^-$ can be calculated using
\begin{equation}
\frac{\diff\sigma}{\diff m_{\pi\pi}\diff
\cos\theta} = N \int_{E_{min}}^{E_{max}}
\frac{|\bar{M}(E^2,s,\cos\theta)|^2
|\vec{k_3^\ast}||\vec{k_5}|}{128\pi^3 |\vec{k_1}|E^2}\diff E\,,\label{eq.pipimassdistribution}
\end{equation}
where the limits of integration are chosen to be identical to the cuts
used to get the experimental rate~\cite{Wang:2014hta}, $N$ is the normalization factor,
$\vec{k_1}$ and $\vec{k_5}$ represent the three-momenta of $e^\pm$ and $\Phi$ in the
center of mass frame, respectively, and $\vec{k_3^\ast}$ denotes the three-momenta of $\pi^\pm$ in the rest frame of the
$\pi\pi$ system. They are given as
\begin{equation}
|\vec{k_1}|=\frac{E}{2}\,, \quad
|\vec{k_3^\ast}|=\frac{1}{2}\sqrt{s-4m_\pi^2}\,, \quad
|\vec{k_5}|=\frac{1}{2E} \lambda^{1/2}\big(E^2,s,M_\phi^2\big) \,.
\end{equation}

\section{Phenomenological discussion}\label{pheno}

\subsection{Characteristics of singlet and octet contributions}

In this work we perform fits simultaneously taking into account the experimental data of the $\pi\pi$
invariant mass distributions and the helicity angular distributions collected in the $Y(4660)$ region of the $e^+e^- \to \psi(2S)
\pi\pi$ process~\cite{Wang:2014hta}. Using the constraint $h_8=h_1 g_8/g_1$,\footnote{One can construct the Lagrangian using Eq.~\eqref{eq.YComponents} as the interpolating field for the $Y(4660)$ directly, which is equivalent to writing the Lagrangian in the form of Eq.~\eqref{LagrangianYpsi2Spipi} with $g_8/g_1 = h_8/h_1 = b/a.$}
there are five free parameters in our fits: $g_{1}$, $h_{1}$, $g_{8}$, $C_{Y\psi^\prime}^{Z_c}$, $C_{Y\psi^\prime}^{\text{loop}}$, and a normalization factor $N$. The parameters $g_1$ and $h_1$
correspond to the low-energy constants in the $Y\psi^\prime\Phi\Phi$ Lagrangian in Eq.~\eqref{LagrangianYpsi2Spipi} for the SU(3) singlet component of the $Y(4660)$, $g_8$ and $h_8$ are the corresponding parameters for the SU(3) octet component.
To illustrate the effect of the SU(3) octet
component, we perform two kinds of fits. In scheme I
we only consider the SU(3) singlet component, the $Z_c$-exchange terms, and the triangle diagrams, while in scheme II,
the SU(3) octet components are taken into account in addition. For scheme I we find one solution, denoted as Fit~I.
For scheme II we find two solutions, denoted as Fit~IIa and Fit~IIb, respectively.
The coupled-channel FSI is considered in all the fits.

\begin{figure}
\centering
\includegraphics[width=\linewidth]{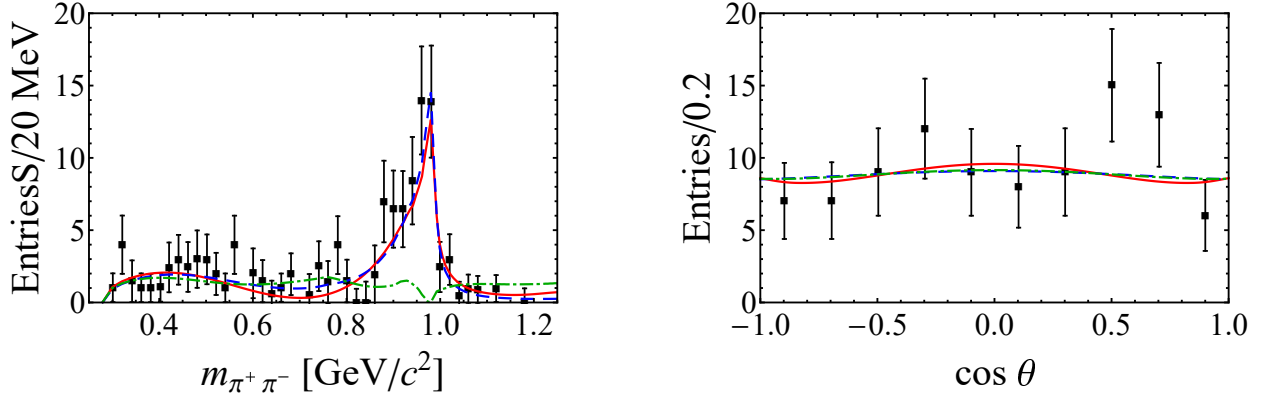}
\caption{Fit results of the $\pi\pi$ invariant mass spectra in the region $4.5 \GeV<E<5.5 \GeV$ (left) and the $\cos\theta$ distribution in the region $4.5 \GeV<E<4.9 \GeV$ (right) in
$e^+e^- \to \psi(2S) \pi^+\pi^-$. Fit I (green dashed) only includes the SU(3) singlet component; Fit IIa (red solid)
and Fit IIb (blue dashed) include
the SU(3) singlet and octet components.
The experimental data are taken from Ref.~\cite{Wang:2014hta}.
}\label{fig.MpipiAndAngular}
\end{figure}

\begin{table}
\caption{\label{tablepar1} The parameters from the fits of
the $\pi\pi$ mass spectrum and the angular distributions in $e^+e^- \to \psi(2S) \pi^+\pi^-$.}
\renewcommand{\arraystretch}{1.2}
\begin{center}
\begin{tabular}{l|ccc}
\toprule
         & Fit~I
         & Fit~IIa
         & Fit~IIb\\
\hline
$g_1~[\text{GeV}^{-1}]$   &    $ 0.94\pm 0.34$  &   $ 0.49\pm 0.21$ &   $ 0.22\pm 0.72$ \\
$h_1~[\text{GeV}^{-1}]$   &    $ 0.53\pm 0.54$  &   $ -1.36\pm 0.34$ &   $ 1.81\pm 0.42$\\
$g_8~[\text{GeV}^{-1}]$     &    0 (\text{fixed})   & $ -1.35\pm 0.45 $  &   $ -0.20\pm 0.60$\\
$C_{Y\psi^\prime}^{Z_c}~[\text{GeV}^{-1}]$     &    $-0.05\pm 0.03$   & $ -0.07\pm 0.04$  &   $ -0.05\pm 0.03$\\
$C_{Y\psi^\prime}^{\text{loop}}~[\text{GeV}^{-1}]$     &    $2.11\pm 0.93$   & $ -0.08\pm 0.26$  &   $ 0.86\pm 0.43$\\
\hline
 ${\chi^2}/{\rm d.o.f.}$ &  $\frac{67.2}{(50-5)}=1.49$  &  $\frac{34.6}{(50-6)}=0.79$   &  $\frac{30.4}{(50-6)}=0.69$    \\
\botrule
\end{tabular}
\end{center}
\renewcommand{\arraystretch}{1.0}
\end{table}

In Fig.~\ref{fig.MpipiAndAngular}, the fitted results of Fits~I, IIa and IIb are shown as the green dot-dashed, red solid, and blue dashed lines, respectively. The fitted parameters as well as the $\chi^2/\text{d.o.f.}$
are given in Table~\ref{tablepar1}.
It is obvious that in Fit I the peak around 1 GeV in the $\pi\pi$ mass spectrum is not reproduced, although the angular distribution
can be described. In contrast, in Fits~IIa and IIb, including the SU(3) octet terms, the fit qualities are improved significantly.
The fit quality of Fit~IIb is a little better than that of Fit~IIa.

\begin{figure}
\centering
\includegraphics[width=\linewidth]{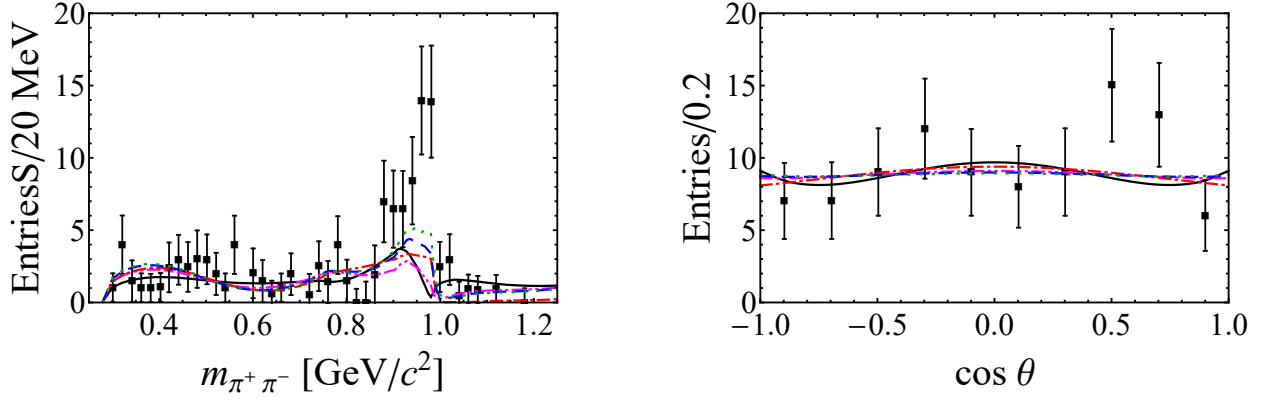}
\caption{Fit results of the $\pi\pi$ invariant mass spectra (left) and the $\cos\theta$ distribution (right) assuming the $Y(4660)$ only contains a SU(3) singlet component and with the setting of $F_K=F_\pi$. The black solid, magenta dash-dot-dotted, red dot-dashed,
blue dashed, and green dotted lines correspond to the results
with the SU(3) breaking parameter $c_A$ fixed at -0.3, 0.15, 0, 0.15, and 0.3, respectively.}\label{fig.MpipiAndAngularSU3breaking}
\end{figure}

Note that in our scheme the SU(3) flavor breaking effect is induced by the corresponding breaking in the
masses of the pseudoscalar meson octet and in the different physical values for the decay constants $F_\pi$ and $F_K$. It is instructive to explore in a more explicit way that whether the SU(3) flavor breaking effect can lead to the experimental dipion invariant mass distributions assuming the $Y(4660)$ only contains the SU(3) singlet component. To account for the different couplings to excite an $s\bar{s}$ pair and a nonstrange pair from vacuum, an SU(3) breaking matrix $X_A=\text{diag}(1,1,1+c_A)$ can be weighted by the SU(3) singlet part in Eq.\eqref{eq.YComponents}. Assuming the $Y(4660)$ only contains the SU(3) singlet component and setting $F_K=F_\pi$, we perform fits with varying $c_A$ in the range of (-0.3,0.3), since the SU(3) breaking effect should be at most around 30\%. In Fig.~\ref{fig.MpipiAndAngularSU3breaking}, the black solid, magenta dash-dot-dotted, red dot-dashed,
blue dashed, and green dotted curves correspond to the fit results with $c_A$ taking values of -0.3, 0.15, 0, 0.15, and 0.3, respectively.
One observes that the sharp peak around $1\GeV$ in the dipion spectra cannot reproduced by the SU(3) breaking effect if the $Y(4660)$ only contains the SU(3) singlet component.

\begin{figure}
\centering
\includegraphics[width=\linewidth]{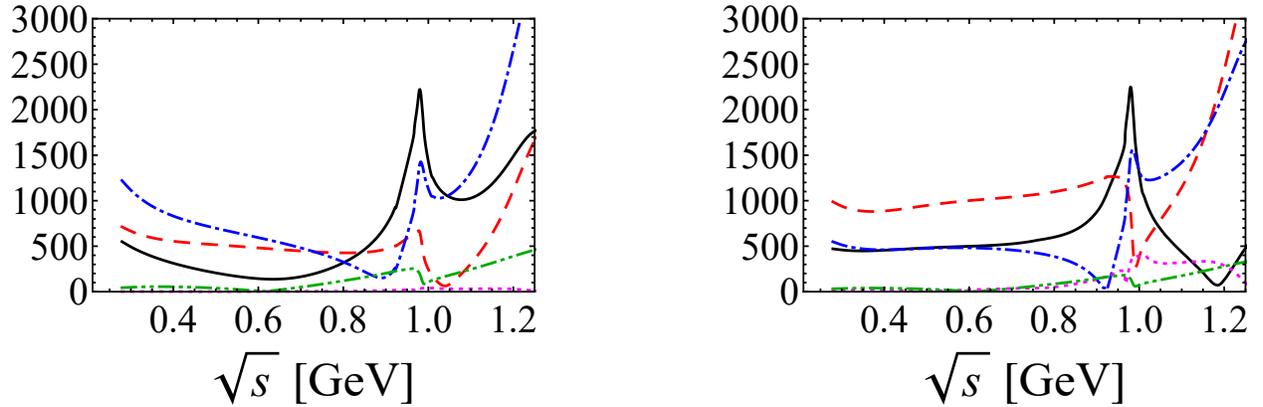}
\caption{The moduli of amplitudes of the chiral contact singlet source, octet source, and the triangle diagrams for $e^+e^- \to \psi(2S) \pi^+\pi^-$ in
Fit~IIa (left) and Fit~IIb (right). The black solid lines represent our best fit results. The red dashed, blue dot-dashed, green dash-dot-dotted, and magenta dotted curves correspond to the contributions of the singlet source, octet source, $Z_c$-exchange, and the triangle diagrams, respectively.}
\label{fig.Moduligihi}
\end{figure}

\begin{figure}
\centering
\includegraphics[width=\linewidth]{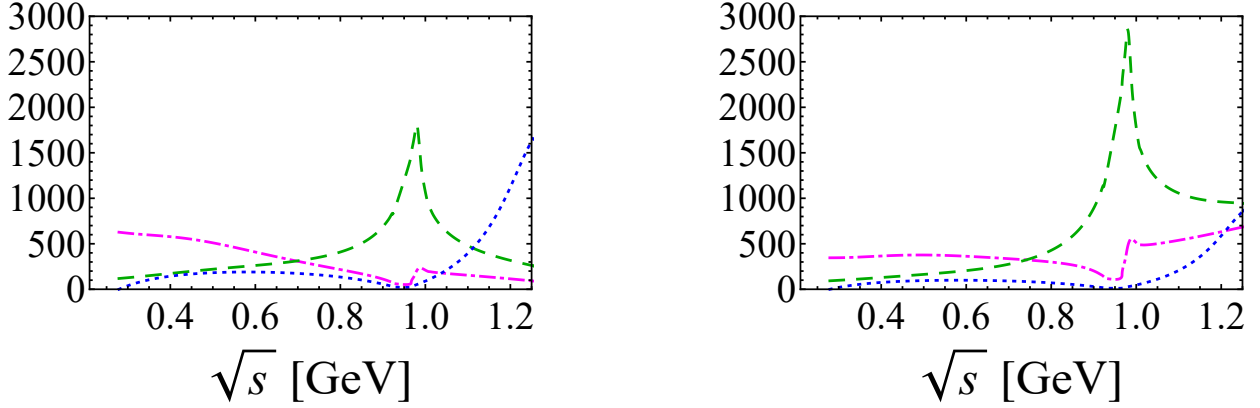}
\caption{The moduli of amplitudes of different transitions for $e^+e^- \to \psi(2S) \pi^+\pi^-$ in
Fit~IIa (left) and Fit~IIb (right). The magenta dot-dashed and blue dotted lines correspond to the $S$- and $D$-wave contributions of the transition via the $Y(4660)\psi(2S)\pi\pi$ contact coupling followed by FSI, i.e., $Y(4660)\to \psi(2S)
\pi\pi \to \psi(2S)
\pi\pi$, respectively, while the green dashed lines represent the $S$-wave contribution of the transition $Y(4660)\to \psi(2S)
K\bar{K} \to \psi(2S)
\pi\pi$.
}
\label{fig.ModulipiK}
\end{figure}

Using the central values of the parameters in Fits~IIa and IIb, we plot the moduli of the amplitudes from different terms.
In Fig.~\ref{fig.Moduligihi}, the black solid lines represent our best-fit result, while the red dashed, blue dot-dashed, green dash-dot-dotted, and magenta dotted curves correspond to the contributions from the chiral contact singlet, octet sources, $Z_c$-exchange, and the triangle diagrams, respectively.
One observes that the contribution of the $Z_c$-exchange and the triangle diagrams show a dip and a broad bump around 1.0 GeV, respectively, and they are much smaller than the contributions of the chiral contact terms.
In the dominant chiral contact terms, the basic characteristic structures of the singlet and the octet contributions are different: the singlet spectra display a broad bump below 1 GeV, while the octet spectra show a sharp peak around 1 GeV, corresponding to the $f_0(980)$. Therefore, the SU(3) octet component is indispensable to reproduce the peak structure in the experimental data.
In Fig.~\ref{fig.ModulipiK}, the magenta dot-dashed and blue dotted lines represent the $S$- and $D$-wave contributions of the transition via the $Y(4660)\psi(2S)\pi\pi$ contact coupling followed by FSI, i.e., $Y(4660)\to \psi(2S)
\pi\pi \to \psi(2S)
\pi\pi$, respectively, while the green dashed lines correspond to the $S$-wave contribution of the transition $Y(4660)\to \psi(2S)
K\bar{K} \to \psi(2S)
\pi\pi$. It is found that around 1 GeV the dominant transition in both Fits~IIa and IIb is $Y(4660)\to \psi(2S)
K\bar{K} \to \psi(2S)
\pi\pi$. In the low energy region, the transition with the $S$-wave $Y(4660)\psi(2S)\pi\pi$ contact coupling plays a major role, which accounts for the bump around 0.5 GeV in the $\pi\pi$ invariant mass spectra.

It is instructive
to analyze the ratio of the parameters for the SU(3) octet component relative to those for
the SU(3) singlet component.
Using the results as shown in Table~\ref{tablepar1}, we have $g_8/g_1=-2.8 \pm 1.5$ for Fit~IIa, and $g_8/g_1=-0.9 \pm 3.9$ for Fit~IIb.
Note that assuming the light-quark component of the $Y(4660)$ is pure $|s \bar{s}\rangle = (
V_1^{\text{light}} -\sqrt{2}V_8^{\text{light}})/\sqrt{3}$ (e.g., in the strange charm- and anticharm-mesons molecule scenario or the $[cs][\bar{c}\bar{s}]$ four-quark scenario), where the definitions of
the singlet and octet components $V_1^{\text{light}}$
and $V_8^{\text{light}}$ have been given below
Eq.~\eqref{eq.YComponents}, the ratio is $-\sqrt{2}$. In the assumption that the light-quark component of the $Y(4660)$ is pure $|u \bar{u}+d\bar{d}\rangle/\sqrt{2} = (\sqrt{2}
V_1^{\text{light}} +V_8^{\text{light}})/\sqrt{3}$ (e.g., in the nonstrange charm- and anticharm-mesons molecule scenario or the $([cu][\bar{c}\bar{u}]+[cd][\bar{c}\bar{d}])\rangle/\sqrt{2}$ four-quark scenario), the ratio is $1/\sqrt{2}$.
Certainly the result of Fit~IIa (values of $g_8/g_1$) differs from the results of the pure $|s \bar{s}\rangle$
or pure $|u \bar{u}+d\bar{d}\rangle/\sqrt{2}$ light-quark component scenarios.
The result of Fit~IIb carries large uncertainty, and its central value is close to the pure $|s \bar{s}\rangle$ light-quark component scenario. As shown in Fig.~\ref{fig.MpipiAndAngular}, both Fits~IIa and IIb describe the $\pi\pi$ invariant mass spectra well, while their theoretical predictions of the angular distributions are different. We note that the present data is limited in statistics, and a better distinction of Fits~IIa and IIb requires new measurement data with higher statistics and smaller error bars.

\section{Conclusions}
\label{conclu}

We have used dispersion theory to study the processes
$e^+e^-\to Y(4660) \to \psi(2S) \pi\pi$. In particular, we have analyzed the roles of the
light-quark SU(3) singlet state and SU(3) octet state in this
transition. The strong FSI, especially the coupled-channel FSI in
the $S$-wave, has been considered model independently by using dispersion theory.
Through fitting to the data of the $\pi\pi$
invariant mass spectra and the angular $\cos\theta$ distributions of $e^+e^-\to Y(4660) \to \psi(2S) \pi\pi$,
we find that the light-quark SU(3) octet state plays a
significant role in the $Y(4660)\psi(2S)\pi\pi$ transition, which
indicates that the $Y(4660)$ contains a large light-quark component.
Thus we conclude that the $Y(4660)$ might not be a pure
conventional charmonium state. For the fit scheme considering both the light-quark SU(3) singlet and SU(3) octet components,
we find two solutions, and both solutions reproduce the $\pi\pi$
invariant mass spectra well.
Notice that the present data is limited in statistics, and new measurement data with higher statistics in the future will be helpful to distinguish between these two solutions.

\section*{Acknowledgments}

We are grateful to Hong-Rong Qi for helpful discussions.
This work is supported in part by the Fundamental Research Funds
for the Central Universities under Grant No.~FRF-BR-19-001A, and by the National Natural Science Foundation of China (NSFC) under Grants No.~11975028, and No.~11974043.


\end{document}